\documentclass[aps,pre,showpacs,twocolumn,floats]{revtex4}
\usepackage{amssymb}

\usepackage{amsmath}
\usepackage{graphicx,psfig}
\usepackage{dcolumn}
\usepackage{bm}

\begin{document}

\title{Phonon-induced relaxation of a two-state system in solids}
\author{Jaroslav Albert, E. M. Chudnovsky, and D. A. Garanin}
\affiliation{ \mbox{Physics Department, Lehman College, City
University of New York,} \\ \mbox{250 Bedford Park Boulevard West,
Bronx, New York 10468-1589, U.S.A.}}
\date{\today}

\begin{abstract}
We study phonon-induced relaxation of quantum states of a particle (e.g.,
electron or proton) in a rigid double-well potential in a solid. Relaxation
rate due to Raman two-phonon processes have been computed. We show that in a
two-state limit, symmetry arguments allow one to express these rates in
terms of independently measurable parameters. In general, the two-phonon
processes dominate relaxation at higher temperature. Due to parity effect in
a biased two-state system, their rate can be controlled by the bias.
\end{abstract}
\pacs{03.65.Yz, 66.35.+a, 73.21.Fg} \maketitle


\section{Introduction}

Relaxation and decoherence in a two state system coupled to environment is a
fundamental problem of quantum physics. It was intensively studied in the
past; see, e.g., the review of Leggett \textit{et al.} \cite{Leggett-review}%
. In the last years the interest to this problem has been revived by the
effort to build solid state qubits. Recently, symmetry implications have
been considered for the problem of a particle in a rigid double-well
potential embedded in a solid \cite{Chudnovsky}. It was demonstrated that
symmetry arguments allow one to obtain parameter-free lower bound on the
relaxation of quantum oscillations in a rigid double well, caused by the
elastic environment. One of the arguments is that the double-well potential
formed by the local arrangement of atoms in a solid is defined in the
coordinate frame of that local atomic environment, not in the laboratory
frame. Another argument is that interactions of the tunneling variable with
phonons must be invariant with respect to global translations and rotations.
When these arguments were taken into account, a simple universal result for
the relaxation rate was obtained \cite{Chudnovsky} in terms of measurable
constants of the solid, with no unknown interaction constants.

The above mentioned universal result refers to the low temperature limit
when the relaxation of a two state system is dominated by the decay of the
excited state due to the emission of one phonon. In this paper we extend the
method developed in Ref. \onlinecite{Chudnovsky} to the study of two-phonon
Raman processes in double-well structures \cite
{Fujisawa1,Fujisawa2,Wiel,Ortner,Naber,Graber}. Such processes can dominate
relaxation at higher temperatures \cite{Orbach,Stoof,Brandes}. We will show
that in the temperature range bounded by the level splitting from below and
by the Debye temperature from above, the rate of the Raman process for a
biased rigid double well is given by a universal expression, very much like
the rate of the direct one-phonon process. The Raman rate is proportional to
the seventh power of temperature, while the one-phonon rate is linear on
temperature. Interestingly, however, at small bias, the Raman rate, unlike
the one-phonon rate, is proportional to the square of the bias. Consequently
Raman processes can be switched on and off by controlling the bias. This
universal result, which is a consequence of the parity of quantum states,
must have important implications for solid-state qubits at elevated
temperatures. Indeed, for an electron in a quantum dot, the rate of a direct
one-phonon process is usually small. If the rate of a two-phonon process can
be made small as well, this means that one can eliminate phonons as a
significant source of relaxation and decoherence of the electron states in
solid-state qubits.

\section{Particle and phonons}

Throughout this paper we shall use units where $\hbar =k_{B}=1$ unless
stated otherwise. In the absence of phonons, the Hamiltonian in the
laboratory frame is
\begin{equation}
{\mathcal{H}}_{o}=\frac{\mathbf{p}^{2}}{2m}+{V(\mathbf{r)}}\,,
\label{Free-Hamiltonian}
\end{equation}
where $\mathbf{r}$ is the radius vector, $\mathbf{p}$ is the momentum, and ${%
m}$ is the mass of the particle (e.g., electron). A long-wave phonon
described by the displacement field $\mathbf{u(r)}$ translates the rigid
double well in space. The Hamiltonian of the system (including the free
phonon field) in the laboratory frame becomes
\begin{equation}
{\mathcal{H}}={\frac{\mathbf{p}^{2}}{2m}}+{V(\mathbf{r-u)}}+{\mathcal{H}}%
_{ph}\ .\   \label{Elect-Phonon Hamilt.1}
\end{equation}
Here, ${{\mathcal{H}}_{ph}\ }$ is the Hamiltonian of the free phonon field.
We intend to obtain a Hamiltonian of the form ${{\mathcal{H}}={{\mathcal{H}}%
_{o}}+{{\mathcal{H}}_{ph}}+{\mathcal{H}}_{e-ph}}$ where the last term
describes the interaction of phonons with the electron in the double well
potential. Using the fact that $\mathbf{u}$ is small, one can expand ${V(%
\mathbf{r-u)}}$ in Taylor series to obtain
\begin{equation}
{\mathcal{H}}={\frac{\mathbf{p}^{2}}{2m}}+{V(\mathbf{r)}}+{\mathcal{H}}_{ph}-%
{\frac{\partial {V}}{\partial {r_{i}}}}{u_{i}}+{\frac{1}{2!}}{\frac{\partial
^{2}{V}}{\partial {r_{i}}\partial {r_{j}}}}{u_{i}}{u_{j}}\ +\ldots \
\label{Taylor exp.-Hamil}
\end{equation}
The first three terms form the interaction-free part of the total
Hamiltonian. The rest of the terms containing powers of $\mathbf{u}$
comprise the electron phonon interaction. It is clear that Eq.\ (\ref{Taylor
exp.-Hamil}) requires detailed knowledge of the potential and its
derivatives. One can, however, obtain Eq.\ (\ref{Elect-Phonon Hamilt.1}) by
performing a unitary transformation on Eq.\ (\ref{Free-Hamiltonian}) with
the help of translation operator $\mathcal{R}{=e^{{i}\mathbf{p}\cdot \mathbf{%
u}}}$
\begin{equation}
{\mathcal{H}}={e^{{-i}\mathbf{p}\cdot \mathbf{u}}}\ \mathcal{H}_{o}{e^{{i}%
\mathbf{p}\cdot \mathbf{u}}}+{{\mathcal{H}}_{ph}}\,.\
\end{equation}
This can be expanded for small $\mathbf{u}$ as
\begin{eqnarray}
{\mathcal{H}} &=&{\mathcal{H}}_{o}+{\mathcal{H}}_{ph}\ +  \notag
\label{Baker-Hausdorf} \\
&+&{i[{\mathcal{H}}_{o}\,,{p_{i}}]u_{i}}+{\frac{i^{2}}{2}[[{\mathcal{H}}%
_{o}\,,{p_{i}}],p_{j}]u_{i}u_{j}}+\ldots \ .
\end{eqnarray}
\newline
Working out the commutators brings one back to Eq.\ (\ref{Taylor exp.-Hamil}%
). However, the use of Eq.\ (\ref{Baker-Hausdorf}) that we are going to
employ allows one to obtain parameter-free results solely in terms of the
energy levels of our effective two-state system without knowledge of the
explicit form of $V(\mathbf{r})$.

We consider the case in which the particle, with good accuracy, is localized
near $\mathbf{r}=\pm \mathbf{R}_{o}$, where $\pm \mathbf{R}_{o}$ are the
energy minima of the left or right wells. Without loss of generality we
assume that $\pm \mathbf{R}_{o}=\pm X_{o}\mathbf{e}_{x}$. The localization
length of the state inside each well is small compared to the distance
between the minima of the double-well potential. The bare ground states
(when tunneling is neglected) in the left and right wells, that we denote by
$|\pm X_{o}\rangle $, are approximately orthonormal,
\begin{equation}
\langle \pm X_{o}|\pm X_{o}\rangle =1\,,\quad \langle -X_{o}|X_{o}\rangle
=0\,.
\end{equation}
The tunneling between the wells leads to the hybridization of the states
given by orthonormal wave functions
\begin{equation}
\left| \psi _{\pm }\right\rangle =\frac{1}{\sqrt{2}}(C_{\pm }\left| {X}%
_{o}\right\rangle \mp C_{\mp }\left| {-X}_{o}\right\rangle )
\label{Eigenstates - explicit form}
\end{equation}
where
\begin{equation}
C_{\pm }=\sqrt{1\pm \varepsilon /\Delta }\,,\qquad \Delta =\sqrt{\Delta
_{o}^{2}+\varepsilon ^{2}}  \label{Coefficients}
\end{equation}
with $\Delta _{o}$ being the tunnel splitting in the unbiased double well
and $\varepsilon $ being the energy bias between the wells. Note that the
double well also has states, $|\psi _{\xi }\rangle $, with energies, $E_{\xi
}$, other then $E_{\pm }$ corresponding to $\left| \psi _{\pm }\right\rangle
$. The energy splitting
\begin{equation}
\Delta =E_{+}-E_{-}
\end{equation}
is considered small compared to the distance from $E_{\pm }$ to other $%
E_{\xi }$. As we shall see, in this limit the summation over all states $%
|\psi _{\xi }\rangle $ renders result for phonon-induced transitions between
$\left| \psi _{\pm }\right\rangle $ that is insensitive to the explicit form
of the potential.

Below we shall deal with the matrix elements of operators $p\equiv p_{x}$, $%
x $, and their combinations. Other components of $\mathbf{p}$ and $\mathbf{r}
$ are irrelevant. Localization of $\left| \psi _{\pm }\right\rangle $ allows
one to compute matrix elements of powers of the operator $x$ with the help
of the relation
\begin{equation}
x|\pm X_{o}\rangle =\pm X_{o}|\pm X_{o}\rangle \,.
\end{equation}
This gives
\begin{eqnarray}
&&\langle \psi _{\pm }|x|\psi _{\pm }\rangle =X_{o}\frac{1}{2}%
(C_{+}^{2}-C_{-}^{2})=X_{o}({\varepsilon }/{\Delta })  \notag \\
&&\langle \psi _{-}|x|\psi _{+}\rangle =X_{o}C_{+}C_{-}=X_{o}({\Delta }_{o}/{%
\Delta })  \notag \\
&&\langle \psi _{\pm }|x^{2}|\psi _{\pm }\rangle =X_{o}^{2}\,,\quad \langle
\psi _{-}|x^{2}|\psi _{+}\rangle =0\,.
\end{eqnarray}
To compute other matrix elements we shall use relations
\begin{equation}
p=m\dot{x}=-im[x,{{\mathcal{H}}_{o}}]  \label{Trick - result}
\end{equation}
and
\begin{equation}
px+xp=m(\dot{x}x+x\dot{x})=m\frac{dx^{2}}{dt}=-im[x^{2},{{\mathcal{H}}_{o}}%
]\,.
\end{equation}
This gives
\begin{eqnarray}
&&\langle \psi _{\xi }|p|\psi _{\xi ^{\prime }}\rangle =im(E_{\xi }-E_{\xi
^{\prime }})\langle \psi _{\xi }|x|\psi _{\xi ^{\prime }}\rangle  \notag \\
&&\langle \psi _{\xi }|p\,x+xp|\psi _{\xi ^{\prime }}\rangle =im(E_{\xi
}-E_{\xi ^{\prime }})\langle \psi _{\xi }|x^{2}|\psi _{\xi ^{\prime }}\rangle
\label{pMEs}
\end{eqnarray}
and thus, $\langle \psi _{-}|px|\psi _{+}\rangle =0$.

As we shall see, perturbation theory for Raman processes requires
computation of the sum
\begin{equation}
\Sigma ={\sum_{\xi \neq +}}\frac{\langle \psi _{-}|p|\psi _{\xi }\rangle
\langle \psi _{\xi }|p|\psi _{+}\rangle }{E_{\xi }-E_{+}}\,.
\label{SigmaDef}
\end{equation}
Application of Eq.\ (\ref{Trick - result}) eliminates the denominator and
yields
\begin{equation}
\Sigma =im{\sum_{\xi \neq +}}\langle \psi _{-}|p|\psi _{\xi }\rangle \langle
\psi _{\xi }|x|\psi _{+}\rangle \,.
\end{equation}
Using the completeness of $|\psi _{\xi }\rangle $ we obtain
\begin{equation}
\Sigma =im[\langle \psi _{-}|px|\psi _{+}\rangle -\langle \psi _{-}|p|\psi
_{+}\rangle \langle \psi _{+}|x|\psi _{+}\rangle ]\,.
\end{equation}
Finally, with the help of the above relations for matrix elements of $x$, $p$%
, and $px$ we get
\begin{equation}
\Sigma =-m^{2}X_{o}^{2}(\Delta _{o}/\Delta )\varepsilon \,.  \label{S}
\end{equation}
This is a mechanism of elimination of unspecified energy levels $E_{\xi }$
from the problem, leading to a universal result.

\section{Raman matrix element}

We are interested in the transition rate between the eigenstates of ${{%
\mathcal{H}}_{o}+{\mathcal{H}}_{ph}\ }$
\begin{equation}
\left| \Psi _{\pm }\right\rangle =\left| \psi _{\pm }\right\rangle \otimes
\left| \phi _{\pm }\right\rangle .\   \label{Eigenstates}
\end{equation}
Here, ${\left| \psi _{\pm }\right\rangle }$ are the tunnel split states of
the double well given by Eq. (\ref{Eigenstates - explicit form}).\ The
states ${\left| \phi _{\pm }\right\rangle }$ are the eigenstates of ${{%
\mathcal{H}}_{ph}\ }$ with energies ${{E_{ph,\pm }}}$. Our goal is to study
Raman scattering processes involving absorption of a phonon of frequency ${%
\omega _{\mathbf{k}}}$ and emission of a phonon of frequency ${\omega _{%
\mathbf{q}}}={\omega _{\mathbf{k}}+\Delta }$, accompanied by the transition
of the particle $\left| \psi _{+}\right\rangle \rightarrow \left| \psi
_{-}\right\rangle ,$ see Fig.1. 
\begin{figure}[tbp]
\unitlength1cm
\begin{picture}(18,3.5)
\centerline{\psfig{file=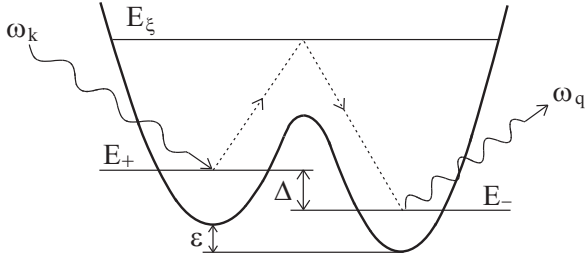,width=7.7cm}}
\end{picture}
\caption{Raman process on tunnel-split levels in a double-well potential,
including virtual transitions to higher levels $E_{\protect\xi }$ [the first
term in Eq. (\ref{Matrix element 1+1}) shown]. }
\label{DoubleWell}
\end{figure}
The Raman rate can be computed with the help of the Fermi golden rule in the
second order in the interaction. The matrix element for this process is the
sum of two matrix elements,
\begin{equation}
{M=M^{(2)}+M^{(1+1)}}.  \label{MSum21p1}
\end{equation}
The first term denotes the first order perturbation on
\begin{equation}
\mathcal{H}_{e-ph}^{(2)}=-\frac{1}{2}\ {[}[\mathcal{H}{_{o}},{p}],{p}{%
]u_{x}^{2}},\   \label{Second order El-Ph Hamil}
\end{equation}
while the second term stands for the second order perturbation on
\begin{equation}
\mathcal{H}_{e-ph}^{(1)}={i}[\mathcal{H}{_{o}},{p}]{u_{x}}.
\label{First order El-Ph Hamil}
\end{equation}
The explicit expressions for ${M^{(2)}}$ and ${M^{(1+1)}}$ are \cite{BHl}
\newline
\begin{equation}
M^{(2)}=\left\langle \Psi _{-}\right| \mathcal{H}_{e-ph}^{(2)}\left| \Psi
_{+}\right\rangle \   \label{Matrix element 2}
\end{equation}
and
\begin{eqnarray}
M^{(1+1)}&=&\sum_{\xi }\ {\frac{\left\langle \Psi _{-}\right| \mathcal{H}%
_{e-ph}^{(1)}\left| \Psi _{\xi }\right\rangle \left\langle \Psi _{\xi
}\right| \mathcal{H}_{e-ph}^{(1)}\left| \Psi _{+}\right\rangle }{E_{+}+\hbar
\omega _{\mathbf{k}}-E_{\xi }}}  \notag \\
&+&\sum_{\xi }\ {\frac{\left\langle \Psi _{-}\right| \mathcal{H}%
_{e-ph}^{(1)}\left| \Psi _{\xi }\right\rangle \left\langle \Psi _{\xi
}\right| \mathcal{H}_{e-ph}^{(1)}\left| \Psi _{+}\right\rangle }{E_{+}-\hbar
\omega _{\mathbf{q}}-E_{\xi }}}.\   \label{Matrix element 1+1}
\end{eqnarray}
Here, ${\left| \Psi _{\xi }\right\rangle }$ is a direct product of the
eigenstates of ${{\mathcal{H}}_{o}\ }$ with the phonon states ${\left| n_{%
\mathbf{k}}-1,n_{\mathbf{q}}\right\rangle }$ in the first term and ${\left|
n_{\mathbf{k}},n_{\mathbf{q}}+1\right\rangle }$ in the second term.

First, we calculate the phonon parts of ${M^{(2)}}$ and $M^{(1+1)}$ using
canonical quantization of the phonon displacements \cite{Kittel}
\begin{equation}
\mathbf{u}=\sqrt{\frac{1}{2\rho V}}\sum_{\mathbf{k}\lambda }\frac{\mathbf{e}%
_{\mathbf{k}\lambda }e^{i\mathbf{k\cdot r}}}{\sqrt{\omega _{\mathbf{k}%
\lambda }}}\left( a_{\mathbf{k}\lambda }+a_{-\mathbf{k,}\lambda }^{\dagger
}\right) ,\   \label{Displacement vector}
\end{equation}
where $\rho $ is the density of the solid, $V$ is its volume, $\mathbf{%
e_{k\lambda }}$ are unit polarization vectors, ${\lambda =t_{1},t_{2},l}$
denotes polarizations, and ${\omega _{\mathbf{k}\lambda }=\upsilon _{\lambda
}k}$ is the phonon frequency that we will usually write as $\omega _{\mathbf{%
k}}$. Writing ${\left| \Psi _{+}\right\rangle =\left| n_{\mathbf{k}},n_{%
\mathbf{q}}\right\rangle \otimes \left| \psi _{+}\right\rangle }$ and ${%
\left| \Psi _{-}\right\rangle =\left| n_{\mathbf{k}}-1,n_{\mathbf{q}%
}+1\right\rangle \otimes \left| \psi _{-}\right\rangle ,}$ for the phonon
matrix element in ${M^{(2)}}$\ one obtains
\begin{eqnarray}
&&\left\langle n_{\mathbf{k}}-1,n_{\mathbf{q}}+1\right| {u_{x}^{2}}\left| n_{%
\mathbf{k}},n_{\mathbf{q}}\right\rangle \   \notag \\
&=&{\frac{1}{\rho V}e_{\mathbf{k}\lambda }^{x}}{e_{\mathbf{q}\epsilon }^{x}}%
e^{i(\mathbf{k}-\mathbf{q})\cdot \mathbf{r}}\sqrt{\frac{{n}_{\mathbf{k}}({n}%
_{\mathbf{q}}+1)}{\omega _{\mathbf{k}}\omega _{\mathbf{q}}}}\equiv M_{ph}.
\label{MPhDef}
\end{eqnarray}
For the phonon matrix elements in $M^{(1+1)}$ one obtains
\begin{eqnarray}
&&\left\langle n_{\mathbf{k}}-1,n_{\mathbf{q}}+1\right| {u_{x}}\left| n_{%
\mathbf{k}}-1,n_{\mathbf{q}}\right\rangle \left\langle n_{\mathbf{k}}-1,n_{%
\mathbf{q}}\right| {u_{x}}\left| n_{\mathbf{k}},n_{\mathbf{q}}\right\rangle
\   \notag \\
&=&\left\langle n_{\mathbf{k}}-1,n_{\mathbf{q}}+1\right| {u_{x}}\left| n_{%
\mathbf{k}},n_{\mathbf{q}}+1\right\rangle \left\langle n_{\mathbf{k}},n_{%
\mathbf{q}}+1\right| {u_{x}}\left| n_{\mathbf{k}},n_{\mathbf{q}}\right\rangle
\notag \\
&=&M_{ph}/2,
\end{eqnarray}
(One can see from the completeness relation that the sum of these two
expressions is $M_{ph}.$)

Next, we evaluate the particle parts of ${M^{(2)}}$ and $M^{(1+1)}.$ For ${%
\left\langle \psi _{-}\right| {[}[{{\mathcal{H}}_{o}\,{p}}]{,p]}\left| \psi
_{+}\right\rangle }$ that enters ${M^{(2)},}$ writing the commutator
explicitly and inserting the identity operator $\mathbb{I}{=\sum_{\xi
}\left| \psi _{\xi }\right\rangle \left\langle \psi _{\xi }\right| }$
results in
\begin{eqnarray}
&&\left\langle \psi _{-}\right| ({\mathcal{H}}_{o}p^{2}-2{p}{\mathcal{H}}_{o}%
{p}+p^{2}{\mathcal{H}}_{o}\ )\left| \psi _{+}\right\rangle   \notag
\label{Matrix elem. with identity} \\
&=&\sum_{\xi }(E_{+}+E_{-}-2E_{\xi })\left\langle \psi _{-}\right| {p}\left|
\psi _{\xi }\right\rangle \left\langle \psi _{\xi }\right| {p}\left| \psi
_{+}\right\rangle .
\end{eqnarray}
The particle part of $M^{(1+1)}$ simpifies to
\begin{eqnarray}
&&-\left\langle \psi _{-}\right| {[{\mathcal{H}}_{o},{p}]}\left| \psi _{\xi
}\right\rangle \left\langle \psi _{\xi }\right| {[{\mathcal{H}}_{o},{p}]}%
\left| \psi _{+}\right\rangle   \notag  \label{remaining matrix elements} \\
&=&(E_{\xi }-E_{-})(E_{\xi }-E_{+})\left\langle \psi _{-}\right| {p}\left|
\psi _{\xi }\right\rangle \left\langle \psi _{\xi }\right| {p}\left| \psi
_{+}\right\rangle .
\end{eqnarray}
Collecting the terms, for $M$ of Eq. (\ref{MSum21p1}) one obtains
\begin{equation}
M=\frac{1}{2}M_{ph}{\sum_{\xi }}\left\langle \psi _{-}\right| {p}\left| \psi
_{\xi }\right\rangle \left\langle \psi _{\xi }\right| {p}\left| \psi
_{+}\right\rangle Q_{\xi },  \label{MFinal}
\end{equation}
where
\begin{eqnarray}
&&Q_{\xi }\equiv E_{+}+E_{-}-2E_{\xi }+(E_{\xi }-E_{-})(E_{\xi }-E_{+})
\notag \\
&&\qquad \times \left[ \frac{1}{E_{\xi }-E_{+}-\omega _{\mathbf{k}}}+\frac{1%
}{E_{\xi }-E_{+}+\omega _{\mathbf{q}}}\right] .
\end{eqnarray}

A common mistake that propagates through literature \cite{AB} is summation
of rates due to $M^{(2)}$ and $M^{(1+1)}$, instead of adding matrix elements
first and then squaring the result and computing the rate. This mistake is
not innocent since $M^{(2)}$ and $M^{(1+1)}$ may cancel leading parts of
each other. Taking into account conservation of energy ${\omega _{\mathbf{q}%
}=\omega _{\mathbf{k}}+\Delta }$ and the relation ${E_{-}=E_{+}-\Delta }$,
one can rewrite this expression as
\begin{equation}
Q_{\xi }=\frac{2\omega _{\mathbf{k}}(\omega _{\mathbf{k}}+\Delta )(E_{\xi
}-E_{+}+\Delta /2)}{(E_{\xi }-E_{+}-\omega _{\mathbf{k}})(E_{\xi
}-E_{+}+\omega _{\mathbf{k}}+\Delta )}.
\end{equation}
One has $Q_{\pm }=\mp \Delta $. We consider the case ${\Delta \ll \omega _{%
\mathbf{k}}}\sim T{\ll \omega _{D},}$ where ${\omega _{D}}$ is the Debye
frequency. Then for $\xi \neq \pm $ the energy differences ${E_{\xi }-E_{+}}$
are considered to be large compared to $\Delta $, that yields
\begin{equation}
Q_{\xi }\cong \tilde{Q}_{\xi }\equiv \frac{2\omega _{\mathbf{k}}(\omega _{%
\mathbf{k}}+\Delta )}{E_{\xi }-E_{+}}.
\end{equation}
From Eq. (\ref{pMEs}) follows that the terms with $\xi =\pm $ in Eq. (\ref
{MFinal}) disappear. Using Eq. (\ref{SigmaDef}), one obtains
\begin{eqnarray}
M &=&\frac{1}{2}M_{ph}{\sum_{\xi \neq +}{}}\left\langle \psi _{-}\right| {p}%
\left| \psi _{\xi }\right\rangle \left\langle \psi _{\xi }\right| {p}\left|
\psi _{+}\right\rangle \tilde{Q}_{\xi }  \notag \\
\qquad  &=&M_{ph}\omega _{\mathbf{k}}(\omega _{\mathbf{k}}+\Delta )\Sigma
\end{eqnarray}
and finally, with the help of Eqs. (\ref{MPhDef}) and (\ref{S}),
\begin{equation}
M=-\frac{1}{\rho V}{e_{\mathbf{k}\lambda }^{x}}{e_{\mathbf{q}\epsilon }^{x}%
\sqrt{\frac{{n}_{\mathbf{k}}({n}_{\mathbf{q}}+1)}{\omega _{\mathbf{k}}\omega
_{\mathbf{q}}}}m^{2}}{X_{o}^{2}}\varepsilon \frac{{\Delta _{o}}}{\Delta }%
\omega _{\mathbf{k}}\left( \omega _{\mathbf{k}}+\Delta \right) .
\end{equation}
Here we have suppressed an irrelevant phase factor$.$ This result for the
Raman matrix element is insensitive to the explicit form of the double-well
potential $V(x).$ It would be a hopeless task to obtain it from Eq. (\ref
{Taylor exp.-Hamil}).

\section{Raman transition rate}

According to the Fermi golden rule \cite{BHl} the Raman rate is given by
\begin{equation}
\Gamma _{2}=\sum_{\lambda \epsilon }\int \frac{d\mathbf{k}d\mathbf{q}}{(2\pi
)^{6}}V^{2}|M|^{2}2\pi \delta (\omega _{\mathbf{q}}-\omega _{\mathbf{k}%
}-\Delta ).\
\end{equation}
The integration variables ${d\mathbf{k}d\mathbf{q}}$ can be written in
spherical coordinates as ${dkdq\ k^{2}q^{2}d\Omega _{\mathbf{k}}d\Omega _{%
\mathbf{q}}}$, and the integral for $T{\ll \omega }_{D}$ can be expressed as
\begin{eqnarray}
\Gamma _{2} &=&\frac{{m^{4}}{X_{o}^{4}\Delta }^{2}{\Delta
_{o}^{2}\varepsilon }^{2}}{2\pi ^{3}\rho ^{2}}B^{2}\int_{0}^{\infty }d{%
\omega _{\mathbf{k}}}d{\omega _{\mathbf{q}}\omega _{\mathbf{k}}}{\omega _{%
\mathbf{q}}}n_{\mathbf{k}}(n_{\mathbf{q}}+1)  \notag \\
&&\times \left( \frac{\omega _{\mathbf{k}}}{\Delta }\right) ^{4}\left[
1+O\left( \frac{\Delta }{\omega _{\mathbf{k}}}\right) \right] \delta (\omega
_{\mathbf{q}}-\omega _{\mathbf{k}}-\Delta ),
\end{eqnarray}
where
\begin{equation}
B\equiv \int \frac{d\Omega _{\mathbf{k}}}{4\pi }\sum_{\lambda }\frac{\left( {%
e_{\mathbf{k}\lambda }^{x}}\right) ^{2}}{{v}_{\lambda }^{3}},  \label{BDef}
\end{equation}
${v_{\lambda }}$ is the velocity of sound with polarization ${\lambda }$,
whereas $n_{\mathbf{k}}=\left( e^{\omega _{\mathbf{k}}/T}-1\right) ^{-1}$ is
the Bose occupation number of a phonon. For $\omega _{\mathbf{k}}\sim T\gg
\Delta $ one obtains the Raman rate
\begin{equation}
\Gamma _{2}=\frac{{m^{4}}{X_{o}}^{4}{\Delta }^{5}{\Delta _{o}^{2}\varepsilon
}^{2}}{2\pi ^{3}\rho ^{2}{\hbar }^{11}}B^{2}\alpha \left( \frac{T}{\Delta }%
\right) ^{7}\left[ 1+O\left( \frac{\Delta }{T}\right) \right] ,
\label{Gamma-R}
\end{equation}
where $\alpha =\int_{0}^{\infty }dxx^{6}e^{x}\left( e^{x}-1\right)
^{-2}=\left( 16/21\right) \pi ^{6}.$ One can see from this integral that the
validity of Eq. (\ref{Gamma-R}) requires at least $T\lesssim {\omega }%
_{D}/10.$ $B$ can be calculated with the help of the transverse-phonons sum
rule $\sum_{t=t_{1},t_{2}}(\mathbf{e}_{\mathbf{k}t}\cdot \mathbf{a})(\mathbf{%
e}_{\mathbf{k}t}\cdot \mathbf{b})=(\mathbf{a}\cdot \mathbf{b})-(\mathbf{a}%
\cdot \mathbf{\hat{k}})(\mathbf{\hat{k}}\cdot \mathbf{b}),$ where $\mathbf{%
\hat{k}\equiv k/}k$, etc. Setting $\mathbf{a=b=}$ ${\mathbf{e}_{x}}$ and
averaging over the directions of $\mathbf{\hat{k}}$ yields
\begin{equation}
B=\int \frac{d\Omega _{\mathbf{k}}}{4\pi }\left( \frac{\hat{k}_{x}^{2}}{{%
v_{l}^{3}}}+\frac{1-\hat{k}_{x}^{2}}{{v_{t}^{3}}}\right) =\frac{2}{3{%
v_{t}^{3}}}+\frac{1}{3{v_{l}^{3}}}.  \label{BFinal}
\end{equation}
According to theory of elasticity \cite{LL-elasticity} $v_{l}>\sqrt{2}v_{t}$%
. Thus, the second term in this expression is a small correction and it can
be neglected. Since we are interested in the region ${\Delta \ll }T$ we can
keep only the leading order on ${T/\Delta }$ in Eq.\ (\ref{Gamma-R}). The
Raman rate for ${\Delta \ll }T{\ll \omega }_{D}$ then becomes
\begin{equation}
\Gamma _{2}=\frac{32\pi ^{3}}{189}\frac{T}{\hbar }\left( \frac{\Delta _{o}}{%
\Delta }\right) ^{2}\left( \frac{\varepsilon }{{\mathcal{E}}_{t}}\right)
^{2}\left( \frac{T}{{\mathcal{E}}_{t}}\right) ^{6}\,,  \label{Raman-final}
\end{equation}
where we have introduced characteristic energy and frequency scales
\begin{equation}
{\mathcal{E}}_{t}=\hbar \Omega _{t}\,,\qquad \Omega _{t}=\left( \frac{\hbar
\rho v_{t}^{3}}{m^{2}X_{o}^{2}}\right) ^{1/4}  \label{Characteristic e&f}
\end{equation}
of the problem that are entirely determined by the parameters of the
unperturbed dot and its elastic environment.

\section{Dot frame calculation}

In this section we will check our result by calculating the Raman rate in
the frame of reference of the dot, as it was done for one-phonon processes
\cite{Chudnovsky}. In the laboratory frame the Lagrangian of the particle is
\begin{equation}
{\mathcal{L}}_{P}=\frac{m}{2}({\Dot{\mathbf{r}}}^{\prime }+\ \Dot{\mathbf{u}}%
)^{2}-\ V(\mathbf{r}^{\prime })\,,  \label{Lab Lagrangian}
\end{equation}
\newline
where $\mathbf{r}^{\prime }$ is the radius vector of the particle of mass $m$
in the coordinate frame rigidly coupled to the double well. The linear
momentum that is canonically conjugated to $\mathbf{r}^{\prime }$ is given
by
\begin{equation}
\mathbf{p}^{\prime }=\frac{\partial {\mathcal{L}}_{P}}{\partial {\Dot{%
\mathbf{r}}}^{\prime }}=m({\Dot{\mathbf{r}}}^{\prime }+\ \Dot{\mathbf{u}})\,.
\end{equation}
The corresponding Hamiltonian is
\begin{equation}
{\mathcal{H}}_{P}(\mathbf{p}^{\prime },\mathbf{r}^{\prime })=\mathbf{p}%
^{\prime }\cdot \mathbf{r}^{\prime }-{\mathcal{L}}_{P}=\frac{\mathbf{p}%
^{\prime 2}}{2m}-\mathbf{p}^{\prime }\cdot \Dot{\mathbf{u}}+V(\mathbf{r}%
^{\prime })\,.
\end{equation}
The full Hamiltonian is ${\mathcal{H}}_{P}+{\mathcal{H}}_{ph}$. Contrary to
the previous model described by Eq.\ (\ref{Taylor exp.-Hamil}), we now have
only one interaction term, $-\mathbf{p}^{\prime }\cdot \Dot{\mathbf{u}}$.
Similarly to Section III, one can write the matrix element for the Raman
processes as
\begin{eqnarray}
M &=&\sum_{\xi }\ {\frac{\left\langle \psi _{-}\right| \mathbf{p}^{^{\prime
}}\cdot \Dot{\mathbf{u}}\left| \psi _{\xi }\right\rangle \left\langle \psi
_{\xi }\right| \mathbf{p}^{^{\prime }}\cdot \Dot{\mathbf{u}}\left| \psi
_{+}\right\rangle }{E_{+}+\hbar \omega _{\mathbf{k}}-E_{\xi }}}  \notag
\label{Matrix element for Dynamical
model} \\
&+&\sum_{\xi }\ {\frac{\left\langle \psi _{-}\right| \mathbf{p}^{^{\prime
}}\cdot \Dot{\mathbf{u}}\left| \psi _{\xi }\right\rangle \left\langle \psi
_{\xi }\right| \mathbf{p}^{^{\prime }}\cdot \Dot{\mathbf{u}}\left| \psi
_{+}\right\rangle }{E_{+}-\hbar \omega _{\mathbf{q}}-E_{\xi }}.}
\end{eqnarray}
Inserting \cite{Kittel}
\begin{equation}
\Dot{\mathbf{u}}=-i\sqrt{\frac{1}{2\rho V}}\sum_{\mathbf{k}\lambda }\mathbf{e%
}_{\mathbf{k}\lambda }e^{i\mathbf{k\cdot r}}\sqrt{\omega _{\mathbf{k}\lambda
}}\left( a_{\mathbf{k}\lambda }-a_{-\mathbf{k}\lambda }^{\dagger }\right)
\end{equation}
into Eq.\ (\ref{Matrix element for Dynamical model}) and evaluating the
matrix elements, we obtain
\begin{equation}
M=\frac{-i}{2}{M}_{ph}\sum_{\xi }\left\langle \psi _{-}\right| p\left| \psi
_{\xi }\right\rangle \left\langle \psi _{\xi }\right| p\left| \psi
_{+}\right\rangle Q_{\xi }\,,
\end{equation}
\newline
which coincides with Eq.\ (\ref{MFinal}) up to an insignificant phase.

\section{Discussion}

We have demonstrated that the two-phonon relaxation of the tunnel-split
states of a particle in a biased solid-state double-well potential can be
expressed in terms of independently measured parameters and without any
unknown constants. Two-phonon processes may dominate relaxation at elevated
temperatures (see below). An interesting observation, however, is that at a
small bias the rate of Eq. (\ref{Raman-final}) is proportional to $%
\varepsilon ^{2}$, while at a large bias it becomes independent of $%
\varepsilon $. This means that one can switch Raman processes on and off by
controlling the bias. This result may seem strange at first, however, it is
a fundamental consequence of quantum mechanics. The reason for this effect
is parity. If we remove the bias, the potential well will become symmetric.
Consequently, the Hamiltonian and the parity operator commute which leads to
eigenstates of even or odd parity. It is easy to see that the states ${%
\left| \psi _{-}\right\rangle \ }$ and ${\left| \psi _{+}\right\rangle \ }$
at ${\varepsilon =0}$ have even and odd parity, respectively. Therefore, the
matrix elements in Eq. (\ref{Matrix element 1+1}) will all vanish.

To find the range of parameters where two-phonon relaxation becomes
important, the rate of the Raman processes should be compared with
one-phonon transition rate \cite{Chudnovsky} that can be written in the form
\begin{equation}
\Gamma _{1}=\frac{\Delta }{3\pi \hbar }\left( \frac{\Delta _{o}}{{\mathcal{E}%
}_{t}}\right) ^{2}\left( \frac{\Delta }{{\mathcal{E}}_{t}}\right) ^{2}\coth
\left( \frac{\Delta }{2T}\right) \,.  \label{Gamma1Ph}
\end{equation}
Notice that Eqs. (\ref{Gamma1Ph}) and (\ref{Gamma1Ph}) do not contain any
unknown interaction parameters. The quantity of interest is the ratio ${%
\Gamma _{2}/\Gamma _{1}}$ which can tell us the importance of the second
order process versus the first order process at various temperatures. If we
take ${T>\Delta }$, the ${\coth (\Delta /2T)}$ in Eq. (\ref{Raman-final})
can be replaced by ${2T/\Delta }$. The above mentioned ratio then yields
\begin{equation}
\frac{\Gamma _{2}}{\Gamma _{1}}=\frac{16}{63}\pi ^{4}\left( \frac{%
\varepsilon }{\Delta }\right) ^{2}\left( \frac{{\mathcal{E}}_{t}}{\Delta }%
\right) ^{2}\left( \frac{T}{{\mathcal{E}}_{t}}\right) ^{6}\,.  \label{ratio}
\end{equation}
\newline
At any given temperature this ratio has a maximum at $\varepsilon =\Delta
_{0}$. For, an electron in a double-well dot with ${X_{o}\sim 10}$ nm
embedded in (or deposited on) a solid with ${\rho \sim 5}$ ${g/cm^{3}}$ and $%
{v_{t}\sim 10^{3}}$ {m/s}, parameter ${\mathcal{E}}_{t}$ is of order $300$
K. Then, for, e.g., $\varepsilon \sim \Delta _{o}\sim 1$ K, Raman processes,
according to Eq.\ (\ref{ratio}), will dominate electron-phonon relaxation
above $33$ K, while below that temperature the relaxation will be dominated
by direct processes. The actual phonon rates for an electron are not likely
to exceed $10^{6}$ s$^{-1}$ even at $T\sim 100$ K. For a proton in a
molecular double well with $X_{o}\sim 0.3$ nm in a solid with ${\rho \sim 5}$
{g/cm}${^{3}}$ and ${v_{t}\sim 10^{3}}$ {m/s}, one gets ${\mathcal{E}}%
_{t}\sim 40$ K. At $\varepsilon \sim \Delta _{o}\sim 1$ mK, according to
Eq.\ (\ref{ratio}), Raman processes will dominate proton-phonon relaxation
above $1$ K, while direct processes will dominate relaxation in the
millikelvin range.

Finally we should note that since our model is based upon bare quantum
states that are well localized in space, it is rigorous for heavy particles,
like, e.g., a proton or an interstitial atom, but is less rigorous for such
a light particle as an electron. Nevertheless, even for an electron our
formulas should provide a good approximation in the limit of weak tunneling
between the wells. Note also that at a large tunnel splitting, the actual
rates for a heavy particle like proton, interstitial atom or defect, can
become so large that the approximation based upon Fermi golden rule will no
longer apply \cite{Comment}. Even in this case, however, the matrix elements
can be expressed in terms of measurable parameters of the quantum well and
the solid.

J. A. thanks Carlos Calero for useful discussions. This research has been
supported by the Department of Energy Grant No. DE-FG02-93ER45487.

\end{document}